# Improving Co-registration for Sentinel-1 SAR and Sentinel-2 Optical images

**Yuanxin Ye [1,*], Chao Yang [1], Bai Zhu [1], Liang Zhou[1], Youquan He [2], and Huarong Jia[3]**

[1] Faculty of Geosciences and Environmental Engineering, Southwest Jiaotong University, Chengdu 610031, China; yc18483685462@my.swjtu.edu.cn (C.Y.); kevin_zhub@my.swjtu.edu.cn (B.Z.); zlup@my.swjtu.edu.cn (L. Z.)
[2] College of Surveying and GeoInformatics, Tongji University, Shanghai 200092, China; 1911211@tongji.edu.cn (Y. H.)
[3] Beijing Institute of Control and Electronic Technology, Beijing 100038, China; 15810687706@163,com (H. J.)
* Correspondence: yeyuanxin@home.swjtu.edu.cn (Y. Y.); Tel.:

**Abstract:** Co-registering the Sentinel-1 SAR and Sentinel-2 optical data of European Space Agency (ESA) is of great importance for many remote sensing applications. However, we find that there are evident misregistration shifts between the Sentinel-1 SAR and Sentinel-2 optical images that are directly downloaded from the official website. To address that, this paper presents a fast and effective registration method for the two types of images. In the proposed method, a block-based scheme is first designed to extract evenly distributed interest points. Then the correspondences are detected by using the similarity of structural features between the SAR and optical images, where the three dimension (3D) phase correlation (PC) is used as the similarity measure for accelerating image matching. Finally, the obtained correspondences are employed to measure the misregistration shifts between the images. Moreover, to eliminate the misregistration, we use some representative geometric transformation models such as polynomial models, projective models, and rational function models for the co-registration of the two types of images, and compare and analyze their registration accuracy under different numbers of control points and different terrains. Six pairs of the Sentinel-1 SAR L1 and Sentinel-2 optical L1C images covering three different terrains are tested in our experiments. Experimental results show that the proposed method can achieve precise correspondences between the images, and the 3rd. Order polynomial achieves the most satisfactory registration results. Its registration accuracy of the flat areas is less than 1.0 10m pixels, and that of the hilly areas is about 1.5 10m pixels, and that of the mountainous areas is between 1.7 and 2.3 10m pixels, which significantly improves the co-registration accuracy of the Sentinel-1 SAR and Sentinel-2 optical images.

## 1. Introduction

Sentinel-1 and 2 are special satellite series of European Copernicus program created by the European Space Agency (ESA). Equipped with new space sensors, Sentinel serial satellites carry out global observation with high revisit rate, and use special wave bands to monitor the Earth surface. At present, Sentinel-1 Synthetic Aperture Radar (SAR) and Sentinel-2 Multiple Spectral Image (MSI) sensors have steadily provided products to the public. The public can download their original data (Level-0) and preprocessed data products (Level-1) free of charge. Although the Sentinel-1 SAR Level-1 (L1) and Sentinel-2 optical Level-1C (L1C) image productions have undergone geometric calibration and rectification, there are still evident misregistration shifts between the images due to their large differences in imaging mechanism. It is necessary to perform further co-registration for the two types of images. This is also a prerequisite step to integrate their image information for the subsequent image processing and analysis tasks such as image fusion [1–4], change detection [5], biological information estimation [6], land cover monitoring [7], classification application [8], etc. Therefore, this research has practical significance in promoting the comprehensive application of the Sentinel SAR and optical products.

Image registration aims to align two or more images acquired by different sensors or on different dates, which mainly include two steps [9]: image matching and geometric correction. Image matching is to detect correspondences or control points (CPs) between images, while geometric correction is to determine a geometric transform using CPs and perform image rectification.

The correspondence detection between the Sentienl-1 SAR and Sentinel-2 optical images belongs to the category of multimodal image matching because they acquire images in different spectral ranges, which results in significant nonlinear radiation or intensity differences between them. In general, multimodal image matching techniques can be divided into feature-based approaches and area-based approaches [10]. In feature-based approaches, salient features such as corners, line intersections, edges, or boundaries of objects are first detected, and then are matched by means of their similarity to find correspondences between images. Currently, Feature-based approaches mainly utilize some invariant feature detectors and descriptors such as Harris [11], SIFT (Scale Invariant Feature Transform) [12], SURF (Speeded Up Robust Features) [13], and their improved versions for image matching [14,15]. However, these features are sensitive to complex nonlinear radiometric changes [16], which makes their matching performance degraded. Area-based methods usually employ a template matching manner to detect correspondences using some similarity measures such as the SSD (Sum of Squared Differences), the NCC (Normalized Cross Correlation), MI (Mutual Information), etc. SSD and NCC are vulnerable to multimodal image matching because they cannot effectively handle nonlinear radiometric differences [10,17]. MI can tolerate radiometric differences to a certain extent [18], but it is difficult to meet the requirement of real time processing due to its large computational cost [17]. Recently, structure and shape features, such as HOG (Histogram of Orientated Gradient) [19], LSS (Local Self-Similarity) [20], HOPC (Histogram of Orientated Phase Congruency) [21] and their improved versions [16,22,23], are integrated as similarity measures for multimodal matching. Since they can capture common properties between multimodal images, they are robust to nonlinear radiometric changes [24]. Accordingly, the area-based methods based on structure and shape features are suitable for the matching of the Sentinel-1 SAR and Sentinel-2 optical images. The main problem of these methods is how to raise their computational efficiency and make the achieved correspondences uniformly distributed over images, which will be addressed in this document.

For remote sensing image rectification, it is a procedure to use rigorous and non-rigorous mathematical models to align the sense images with the reference images. Image rectification aims to find the optimal mathematical transformation model that can fit geometric distortions between images. In general, the precise geometric correction of remote sensing images is performed by using rigorous mathematical models and orbit ephemeris parameters of satellites. However, the orbit ephemeris parameters are not available for ordinary users. Accordingly, non-rigorous or empirical mathematical models are often used for remote sensing image registration. These models mainly include Direct Linear Transformation (DLT), affine transformation models, polynomial models, Rational Function Models (RFMs), projective transformation models, etc. El-Manadili and Novak suggested the use of DLT for geometric correction of SPOT images [25]. Okamoto et al., discussed affine models for SPOT level 1 and 2 stereo scenes and achieves good results [26]. Subsequently, geometric correction based on polynomial and projective models is performed on high resolution images such as SPOT and IKONOS images, and the correction accuracy is evaluated under different number of CPs and different terrains [27–29]. Also, RFMs are widely applied for geometric correction and 3D reconstruction of high resolution images [30–35]. These research illustrates that the non-rigorous or empirical geometric models can be used for precise correction of remote sensing images. Accordingly, these empirical models will be explored for the co-registration of the Sentinel-1 SAR and Sentinel-2 optical images in this document.

Sentinel serial satellites have provided a large number of time serial products since they are put into operation. Many researchers have carried out relevant registration research and geolocation accuracy analysis on the published product data. For example, Schubert et al. and Languille et al. presented first results of geolocation assessment for Sentinel-1A SAR images and Sentinel-2A optical images, respectively [36,37]. Their results show that both the SAR L1 productions and the optical L1C productions meet their geolocation accuracy specifications defined by ESA. Then, the geolocation accuracy of Sentinel-1A and 1B SAR L1 productions are further validated and improved using high precise corner reflectors [38,39] Meanwhile, Yan et al. characterized the misregistration errors and proposed a sub-pixel registration method for Sentinel-2A multi-temporal images [40]. In addition, some researchers also conducted the registration tests between Sentinel-2 optical images and other satellite images such as Landsat-8 images. Barazzetti et al. analyzed the co-registration accuracy of the Sentinel-2 and Landsat-8 optical images [41]. Then, Yan et al. proposed an automatic sub-pixel registration method for the two types of images [42]. Subsequently, Skakun et al. explored a phase correlation method for CP detection, and evaluated a variety of geometric transformation models for the co-registration of Sentinel-2 and Landsat-8 optical images [43]. Recently, Stumpf et al. improved the registration accuracy of Sentinel-2 and Landsat-8 optical images for Earth surface motion measurements by a dense matching method [44].

When we use the Sentinel-1 SAR L1 images and Sentinel-2 optical L1C images which are directly downloaded from the official website[1], it can be observed that the two types of images have significant misregistration shifts. This is because the Sentinel-1 SAR images have not undergone the terrain correction, and the side-looking geometry of the SAR images results in significant geometric distortions with respect to Sentinel-2 the optical images. Current methods main perform the co-registration of the Sentinel-1 SAR and Sentinel-2 optical images by using the Sentinel Application Platform (SNAP) toolbox[2] to carry out the terrain correction process for Sentinel-1 SAR images [1–3], and assume that such process can align the two types of images well [1]. However, the registration accuracy is quite satisfactory because the data instructions given by ESA [45,46] illustrate that the Sentienl-1 SAR L1 productions and the Sentinel-2 optical L1C productions have a geolocation accuracy of 7m ($3\sigma$) and 12m ($3\sigma$), respectively. In other words, the two types of images may have a registration error of about 2 10m pixels. Moreover, the terrain correction process is quite time consuming which cannot meet real-time applications. To address that, this paper presents a fast and robust matching method for the directly downloaded Sentinel-1 and Sentinel-2 images , which is used to measure their misregistration shifts and determine an optimum geometric transformation model for their co-registration. Firstly, we propose a block-based image matching scheme based on structural and shape features. In this procedure, a blocked extraction strategy with the Features from Accelerated Segment Test (FAST) operator is designed to detect evenly distributed interest points from images. Then structure features are extracted by using gradient information of images, and the three dimension (3D) phase correlation (PC) is used as the similarity measure for detecting correspondences by a template matching manner. The proposed matching scheme could insure the uniform distribution of correspondences by the block extraction strategy, handle nonlinear radiometric differences using the similarity of structure features, and accelerate image matching because of the use of 3D PC. Subsequently, these correspondences are used to measure and analyze the misregistration shifts between the Sentinel SAR and optical images. Meantime, we also compare the registration accuracy of the current representative geometric transformation models (such as polynomials, projective transformations and RFMs), and analyze various acquisition factors that influence their performance (e.g. terrain, number of CPs, etc.). Finally, we determine an optimal geometric transformation model for the co-registration of the Sentinel SAR and optical images, and improve the registration accuracy of such images. The main contributions of this paper are as follows.

(1) We design a block-based matching scheme based on structural features to detect evenly distributed correspondences between the Sentinel-1 SAR L1 images and the Sentinel-2 optical L1C images, which is computationally effective and is robust to nonlinear radiometric differences.
(2) We precisely measure the misregistration shifts between the Sentinel SAR and optical images.

---

[1] https://scihub.copernicus.eu/

[2] The SNAP Toolbox can be downloaded from this website (http://step.esa.int/main/toolboxes/snap/)

(3) We compare and analyze the registration accuracy of various geometric transformation model, and determine an optimal model for the co-registration of the Sentinel SAR and optical images.

This paper is structured as follows. Section 2 first gives the introduction for the Sentinel-1 SAR and Sentinel-2 optical image productions, and then Section 3 describes the proposed co-registration method for the two types of images. Subsequently, quantitative experimental results are presented in Section 4. Finally, we conclude with a discussion of the results and a recommendation for future work.

**2. Sentinel-1 SAR and Sentinel-2 optical data introduction**

The Sentinel-1 SAR instrument provides a long-term Earth observation with high reliability and improved revisit time at C-band, which captures data in four operation modes [47]: Stripmap (SM), Interferometric Wide swath (IW), Extra Wide swath (EW), and Wave (WV). The IW mode is the primary acquisition mode over land and satisfies the majority of service requirements. It provides a large swath of 250km at $5 \times 20$m spatial resolution (single look). The IW L1 productions include the Single Look Complex (SLC) data and the Ground Range Detected (GRD) data. The IW L1 GRD data are the multi-looked intensity or gray images, which have the similar visual features with optical images and are available in two resolutions ($20 \times 22$m and $88 \times 87$m). The data in the two resolution are respectively resampled at the pixel spacing of $10 \times 10$m and $40 \times 40$m, and then they are distributed to users. In this paper, the 10m data are used as it provides more spatial detail than the 40m data. The Sentinel-1 IW L1 GRD 10m productions are provided in geolocated tiles of about 25000 $\times$ 16000 pixels in the geographic longitude/latitude coordinate in the World Geodetic System 84 (WGS84) datum. The downloaded Sentinel-1 L1 GRD images have not undergone the terrain correction, which results in they have significant geometric distortions related to their side-looking geometry.

The Sentinel-2 carries the MSI that acquires data at 13 spectral bands: 4 visible and near-infrared bands with 10m, 6 near-infrared, red edge and short wave infrared bands with 20m, and 3 bands with 60m resolution. Similar to Sentinel-1, the Sentinel-2 10m bands are used in this paper because of its higher resolution compared with the other bands. The Sentinel-2 MSI covers a field of view providing a 290km swath and performs a systematic global observation with high revisit frequency. The processed L1C 10m productions are available in geolocated tiles of $10980 \times 10980$ pixels with the UTM projection in the WGS84 datum.

According to the above description, the Sentinel-1 IW L1 GRD 10m productions and the Sentinel-2 L1C optical 10m productions present fine image details and have the same resolution, which makes the two types of images easily intergraded for the following remote sensing image application. Therefore, this paper carries out the co-registration between the Sentinel-1 SAR L1 IW GRD 10 images and the Sentinel-2 optical L1C 10m images.

**3. Methodology**

*3.1. Detection of correspondences or CPs*

The first step of registering the Sentinel SAR and optical images is to achieve reliable and precise correspondences or CPs between such images. Due to different imaging modes, The Sentinel SAR and optical images have significant nonlinear radiometric differences, that is, the images present quite different intensity and texture information. To address that, we employ structure features for correspondence detection because structure properties can be preserved between SAR and optical images in spite of significant radiometric differences [16,21,48]. Moreover, to rapidly extract evenly distributed correspondences, this paper designs a block-based matching strategy and perform image matching in frequency domain for acceleration. Meanwhile, this matching scheme also fully considers the characteristics of Sentinel images such as large data volume (more than 10000*10000 pixel) and initial georeference information. Generally speaking, the proposed matching method consists of four steps: interest point detection, structure feature extraction, interest point matching, and mismatch elimination, where the Sentinel-2 optical image is used as the reference image and the Sentinel-1 SAR image is used as the sensed image.

3.1.1 Interest point detection

Currently, there are many classical operators to detect interest points such as Moravec [49], Harris [11], Difference of Gaussians (DoG) [12], the Features from Accelerated Segment Test (FAST) [50] and so on. Given the decisive advantage that is its faster high computation speed than others, the FAST operator, therefore, is employed to detect interest points in this paper.

In order to detect evenly distributed interest points in the reference image, we design a block detection strategy on based the FAST operator, which named the block-based FAST operator. Specifically, the image is first divided into $N \times N$ non-overlapping blocks, and the FAST value is computed for each pixel of every block. Then $K$ pixel points with the largest FAST values are selected as the interest points in a block. As a result, we can obtain $N \times N \times K$ interest points in the image, where the values of $N$ and $K$ are decided by users. From Figure 1, it's obvious to find that the block-based FAST operator has a better representation than the initial one in extracting evenly distributed interest points, where $N$ and $K$ are set to 20 and 1, respectively.

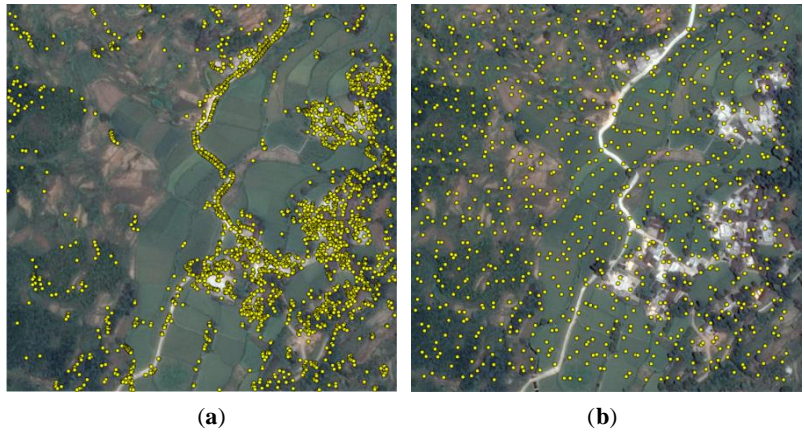

**Figure 1.** Extraction results of interest points by two different strategies. (**a**) Initial FAST operator. (**b**) Block-based FAST operator.

**3.**1.2 Structure feature extraction

Once a number of interest points are detected in the reference image, we determine a template window centered on an interest point, and predict its corresponding search window in the sensed image by the initial georeference information. Then we extract structure features for both the template and search windows, and use them for correspondence detection by a template matching manner. The selection of structural feature descriptor is crucial for image matching, and it needs to consider both the robustness and computational efficiency. Accordingly, a recently published structural feature descriptor [24] named CFOG (Channel features of Orientated Gradients), is employed for correspondence detection because it presents faster and more robust matching performance than other state of the art structural feature descriptors such as HOG, LSS and HOPC.

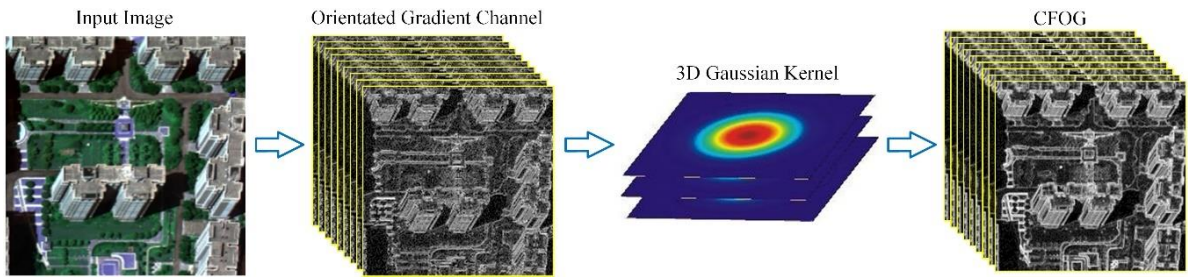

**Figure 2.** The construction process of CFOG.

CFOG is built based on dense orientated gradient features of images, and its construction process (see Figure 2) mainly including two part: orientated gradient channels and 3D Gaussian convolution. Given an image, the $m$ orientated gradient channels of the image first are calculated. The $g_i$ is used to represent each orientated gradient channel $1 \leq i \leq m$. In the actual calculation, the gradient amplitudes in the horizontal and vertical directions are used to calculate the value of each orientated gradient channel, which are denoted as $g_x$ and $g_y$, respectively, which can be expressed by the following equation:

$$g_\theta = abs(\cos\theta \bullet g_x + \sin\theta \bullet g_y) \qquad (1)$$

Where, $\theta$ represents the orientation of the gradient, $abs$ represents the absolute value, and its purpose is to limit the gradient direction within the range of 0° to 180°. After that, the feature channel of convolution is realized by using a 3D-like Gaussian kernel.

$$g_\theta^\sigma = G_\sigma \otimes g_\theta \qquad (2)$$

Where, $\otimes$ denotes the convolution operation, $G$ denotes the Gaussian function, and $\sigma$ is its standard deviation. The aim of such operation is to integrate orientated gradient information of neighborhood for feature description, which can improve the robustness to noise and nonlinear radiometric differences.

3.1.3 Interest point matching

CFOG yields the 3D pixel-wise descriptors with large data volume for both the template and search windows. It is usually time consuming if using these descriptors to perform the template matching by the traditional similarity measures (e.g., SSD and NCC) in spatial domain. Accordingly, we carry out the similarity evaluation in frequency domain, where the 3D PC is used as the similarity metric. The 3D PC can effectively speed up the image matching by using the Fast Fourier Transform (FFT) because the dot product operation in frequency domain is equal to the convolution operation in spatial domain. The details of the matching technique are as follows.

Given the CFOG descriptors of the template and search windows, their relationship is expressed as follows:

$$t(x, y, z) = s(x - x_0, y - y_0, z) \qquad (3)$$

Where $t(x, y, z)$ represents the CFOG descriptors of the template window, $s(x, y, z)$ represents the CFOG descriptors of the search window, $(x_0, y_0)$ is the offset of the two windows and z is the dimensionality of CFOG. Let's use $T(u, v, w)$ and $S(u, v, w)$ to represent the corresponding $t(x, y, z)$ and $s(x, y, z)$ in frequency domain. According to the translation property of Fourier transform, the corresponding relation can be expressed as:

$$T(u, v, w) = S(u, v, w) e^{-i(ux_0 + vy_0)} \tilde{\gamma} \qquad (4)$$

Their normalized cross-power spectrum can be expressed as:

$$\frac{T(u, v, w) S(u, v, w)^*}{\left| T(u, v, w) S(u, v, w)^* \right|} = e^{-i(ux_0 + vy_0)} \tilde{\gamma} \qquad (5)$$

In the above equation, * denotes the complex conjugate. According to the translation theory, a Dirac delta function $\delta(x - x_0, y - y_0)$ can be obtained by the inverse Fourier transform of the normalized cross-power spectrum. The function has an obvious sharp peak value at the offset position, while the value of other positions is close to zero. Accordingly, the offset $(x_0, y_0)$ between the template and search windows can be determined by the peak position.

3.1.4 Mismatch elimination

Due to some occlusions such as clouds and shadow, it is inevitable to cause some correspondences with large errors during the matching process. Therefore, it is necessary to eliminate these mismatches to improve the matching quality. In this paper, the Random Sample Consensus (RANSAC) algorithm is applied to remove these outliers [51], achieving the final correspondences between the Sentinel SAR and optical images. After that, these correct correspondences are used for the subsequent analysis of geometric transformation models.

*3.2. Mathematical models of geometric transformation*

After sufficient CPs are obtained between the Sentinel SAR and optical images, it is crucial to determine an optimum geometric transformation model for the co-registration of such images. Accordingly, we investigate several representative transformation models, and analyze their registration accuracy. These models include polynomial models, projective models, and RFMs, which are described in details as follows.

3.2.1. Polynomial models

Image correction based on polynomials does not take into account the geometric process of spatial imaging, and deals with image deformation by means of a mathematical simulation. Geometric distortions of remote sensing images are often quite complicated because they are caused by a variety of factors such as imaging models, Earth curvature, atmospheric refraction, etc. In general, these distortions can be considered as the combined effect of translation, rotation, scaling, deflection, bending, affine and higher-order based deformation. It is difficult to build a rigorous mathematical model to correct these distortions, whereas an appropriate polynomial can be used to fit the geometric relationship between images. Due to their simplicity and availability, polynomials have been widely applied for image correction by many commercial software packages. Common polynomials include the 1st. Order polynomial, the 2nd. Order polynomials, the 3rd. Order polynomials and higher-order polynomials. The general two-dimensional polynomial model is expressed as follows:

$$x = \sum_{i=0}^{m}\sum_{j=0}^{n} a_{ij} X^i Y^j$$
$$y = \sum_{i=0}^{m}\sum_{j=0}^{n} b_{ij} X^i Y^j \quad (6)$$

Where $(x, y)$ are the coordinates of the sensed image, $(X, Y)$ are the coordinates of the reference image, and $(a_{ij}, b_{ij})$ are the polynomial parameters. The minimum number $N$ of correspondences for solving the parameters is related to the order $n$ of polynomials, which is calculated by

$$N = \frac{(n+1)(n+2)}{2} \quad (7)$$

3.2.2. Projective models

The projective transformation means that if a straight line in one image is mapped to another image, it is still a straight line, but the parallel relation between lines cannot be maintained. Two-dimensional plane projective model is a linear transformation of homogeneous three-dimensional vectors. Under homogeneous coordinate system, the projective transformation can be described in the following nonsingular 3x3 matrix form.

$$\begin{pmatrix} x'_1 \\ x'_2 \\ x'_3 \end{pmatrix} = \begin{pmatrix} h_{11} & h_{12} & h_{13} \\ h_{21} & h_{22} & h_{23} \\ h_{31} & h_{32} & 1 \end{pmatrix} \begin{pmatrix} x_1 \\ x_2 \\ x_3 \end{pmatrix} \quad (8)$$

By setting $u_1 = x'_1/x'_3$ and $u_2 = x'_2/x'_3$, Equation (8) is rewritten as

$$u_1 = \frac{h_{13} + h_{11}x_1 + h_{12}x_2}{1 + h_{31}x_1 + h_{32}x_2}$$
$$u_2 = \frac{h_{23} + h_{21}x_1 + h_{22}x_2}{1 + h_{31}x_1 + h_{32}x_2} \tag{9}$$

Where $(u_1, u_2)$ are the coordinates of the sensed image, and $(x_1, x_2)$ are the coordinates of the reference image.

With the development of geometric transformation models, the projective transformation has been extended by changing the properties of same denominator and increasing the number of parameters, thus forming some flexible projective models for remote sensing image correction. These projective models are as follows.

The projective model of 10 Parameters is

$$x = \frac{a_0 + a_1 X + a_2 Y}{1 + a_3 X + a_4 Y}$$
$$y = \frac{b_0 + b_1 X + b_2 Y}{1 + b_3 X + b_4 Y} \tag{10}$$

The projective model of 22 Parameters is

$$x = \frac{a_0 + a_1 X + a_2 Y + a_3 XY + a_4 X^2 + a_5 Y^2}{1 + a_6 X + a_7 Y + a_8 XY + a_9 X^2 + a_{10} Y^2}$$
$$y = \frac{b_0 + b_1 X + b_2 Y + b_3 XY + b_4 X^2 + b_5 Y^2}{1 + b_6 X + b_7 Y + b_8 XY + b_9 X^2 + b_{10} Y^2} \tag{11}$$

The projective model of 38 Parameters is

$$x = \frac{a_0 + a_1 X + a_2 Y + a_3 XY + \cdots a_8 X^3 + a_9 Y^3}{1 + a_{10} X + a_{11} Y + a_{12} XY + \cdots a_{17} X^3 + a_{18} Y^3}$$
$$y = \frac{b_0 + b_1 X + b_2 Y + b_3 XY + \cdots b_8 X^3 + b_9 Y^3}{1 + b_{10} X + b_{11} Y + b_{12} XY + \cdots b_{17} X^3 + b_{18} Y^3} \tag{12}$$

Where $(x, y)$ are the coordinates of the input image, $(X, Y)$ are the coordinates of the reference images, and $(a_0, a_1, \ldots a_n, b_0, b_1, \ldots, b_n)$ are the model parameters. The minimum number of CPs for the 10-parameter projective model is 5, that for 22-parameter projective model is 11, and that for the 38-parameter projective model is 14.

3.2.3. RFM

Nowadays, the RFM is one of the commonly used geometric transformation models for remote sensing image correction. Lots of research has shown that the RFM can be used as a generalized sensor model, which provides an exact approximation of rigorous sensor models for many satellite images, especially for high-resolution satellite images. This means that the RFM can replace rigorous senor models for precise geometric correction of high-resolution images. Due to its generalization, the RFM has been used as a standard data transfer format for high-resolution images. The RFM relates the object space $(X, Y, Z)$ coordinates to image space $(r, c)$ coordinates in the form of rational functions that are ratios of two polynomials. The generic form of the RFM is given as:

$$r_n = \frac{P_1(X_n, Y_n, Z_n)}{P_2(X_n, Y_n, Z_n)}$$
$$c_n = \frac{P_3(X_n, Y_n, Z_n)}{P_4(X_n, Y_n, Z_n)} \tag{13}$$

Where $(r_n, c_n)$ are the coordinates in image space, $(X_n, Y_n, Z_n)$ are the coordinates in object space, $(P_1, P_2, P_3, P_4)$ denote the polynomials of $(X_n, Y_n, Z_n)$, and their highest order is often limited to 3. In such a case, the polynomial has the following form.

$$P(X_n, Y_n, Z_n) = a_0 + a_1 X_n + a_2 Y_n + a_3 Z_n + a_4 X_n Y_n + a_5 X_n Z_n + a_6 Y_n Z_n a_7 X_n^2 + a_8 Y_n^2 + a_9 Z_n^2 + a_{10} X_n Y_n Z_n + a_{11} X_n^3 + a_{12} Y_n X_n^2 + a_{13} Y_n Z_n^2 + a_{14} X_n Y_n^2 + a_{15} X_n^3 + a_{16} X_n Z_n^2 + a_{17} Y_n^2 Z_n + a_{18} X_n^2 Z_n + a_{19} Z_n^3 \quad (14)$$

Where $(a_0, a_1, ... a_{18}, a_{19})$ are the model parameters called rational polynomial coefficients.

The RFM has two types with same denominator (i.e., $P_2 = P_4$) and different denominator (i.e., $P_2 \neq P_4$). The minimum number of CPs is determined by the order of this model, which is given in Table 1. It is worth noting that the RFM is a 3 three-dimensional polynomial model when $P_2 = P_4 = 1$.

Table 1. Nine cases for the RFM

| Order of Polynomials | Cases | Number of parameters | Min. number of CPs |
|---|---|---|---|
| 1 | $P_2 = P_4 = 1$ | 8 | 4 |
| 1 | $P_2 = P_4$ | 11 | 6 |
| 1 | $P_2 \neq P_4$ | 14 | 7 |
| 2 | $P_2 = P_4 = 1$ | 20 | 10 |
| 2 | $P_2 = P_4$ | 29 | 15 |
| 2 | $P_2 \neq P_4$ | 38 | 19 |
| 3 | $P_2 = P_4 = 1$ | 40 | 20 |
| 3 | $P_2 = P_4$ | 59 | 30 |
| 3 | $P_2 \neq P_4$ | 78 | 39 |

*3.3. Image co-registration*

Based on the above matching approach and geometric transformation models, this paper proposes a technical solution to measure the misregistration shifts and improve the co-registration accuracy between the Sentinel SAR and optical images. Figure 3 shows the technique solution, which includes the following steps.

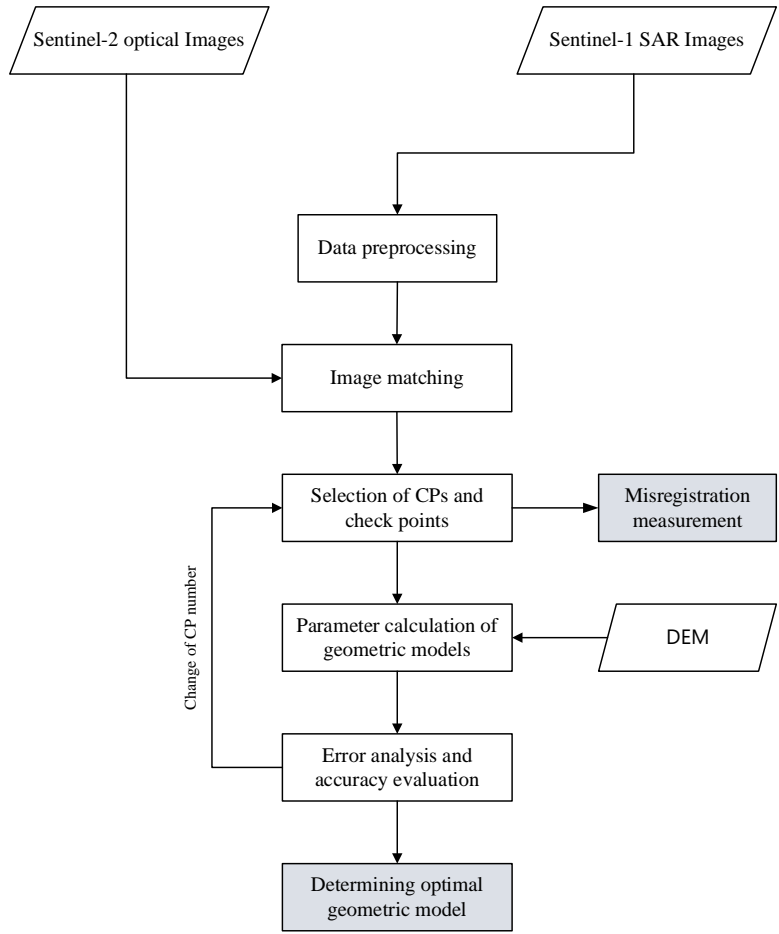

**Figure 3.** Flowchart of misregistration measurement and improvement for the Sentinel-1 SAR and the Sentinel-2 optical images.

3.3.1. Data preprocessing

According to the data description in Section 2, the Sentinel-1 SAR L1 images and Sentinel-2 optical L1C images have different projection coordinate systems and different sizes. Accordingly, a reprojection and a crop operation are carried out for the two types of images before co-registration. In this study, the SAR images are reprojected into the coordinate system (i.e., UTM and WGS84) of the optical images. Considering that the SAR images have some geolocation errors, and their size are larger than the optical images, we crop the SAR images by using the geographical range of the optical images with adding some margin (e.g. 100 pixels) in each direction. Such process ensure that the two types of images can cover the same geographical range.

3.3.2. Image matching

We detect sufficient and evenly distributed correspondences between images by the technique described in Section 3.1. Some of these correspondences are selected as CPs which are used to calculate the parameters of geometric transformation models, while others are used as checkpoints to evaluate the registration accuracy of these models.

3.3.3. Misregistration measurement

The correspondences between the Sentinel SAR and optical images should have the same geographical coordinates if the two types of images are co-registered exactly. Based on this precondition, the differences of geographical coordinates of the correspondences correspond to the misregistration shifts between the Sentinel SAR and optical images. Accordingly, the misregistration shift is quantified as:

$$\Delta x = x_i^s - x_i^o \qquad \Delta \bar{x} = \frac{1}{m}\sum_{i=1}^{m}|\Delta x|$$

$$\Delta y = y_i^s - y_i^o \qquad \Delta \bar{y} = \frac{1}{m}\sum_{i=1}^{m}|\Delta y| \qquad (15)$$

$$\Delta s = \sqrt{\Delta x^2 + \Delta y^2} \qquad \Delta \bar{s} = \frac{1}{m}\sum_{i=1}^{m}\Delta s$$

Where $x_i^s, y_i^s$ and $x_i^o, y_i^o$ are the matched location for the correspondence $i$ between the Sentinel SAR and optical images, $\Delta x, \Delta y$ are the misregistration shifts (units 10m pixels) in $x$ and $y$ directions for one correspondence, and $\Delta \bar{x}, \Delta \bar{y}$ are the mean misregistration shifts in $x$ and $y$ directions. $\Delta \bar{s}$ is the mean misregistration shift between the SAR and optical images, and there are a total of $m$ correspondences.

### 3.3.4. Parameter calculation of geometric models

Based on the obtained CPs, we calculate the parameters of the employed geometric models including polynomial models, projective models, and RFMs using the least square method. Meanwhile, the Digital Elevation Model (DEM) is introduced for the parameter calculation of RFMs.

### 3.3.5. Accuracy analysis and evolution

We adjust the number of CPs to perform the registration experiment, and then compute the root mean square errors (RMSEs, Equation 16) of checkpoints for accuracy evaluation of each geometric model. Finally, the optimal geometric model is determined for the co-registration of the SAR and optical images.

$$RMSE = \sqrt{\frac{\sum_{i=1}^{N}\sqrt{\left(x_i^s - T(x_i^o, y_i^o)\right)^2 + \left(y_i^s - T(x_i^o, y_i^o)\right)^2}}{N}} \qquad (16)$$

Where $T$ denotes the geometric transformation model, and $N$ is the number of checkpoints.

## 4. Experiments

### 4.1. Experimental data

We select six pairs of Sentinel-1 SAR L1 images and Sentinel-2 optical L1C images for the experiment. These images locate in China and Europe (see **Figure** 4), and cover three types of different terrains such as flat areas, hilly areas, and mountainous areas. Each type includes two pairs of images. These images are almost cloud-free, which makes the applied matching technique reliable to detect correspondences between such images. Table 2 gives the detail description of experimental data. In addition, 30m DEM data (from ASTER GDEM) are used as the elevation reference data for image correction. All these experimental data are shown in Figure 5.

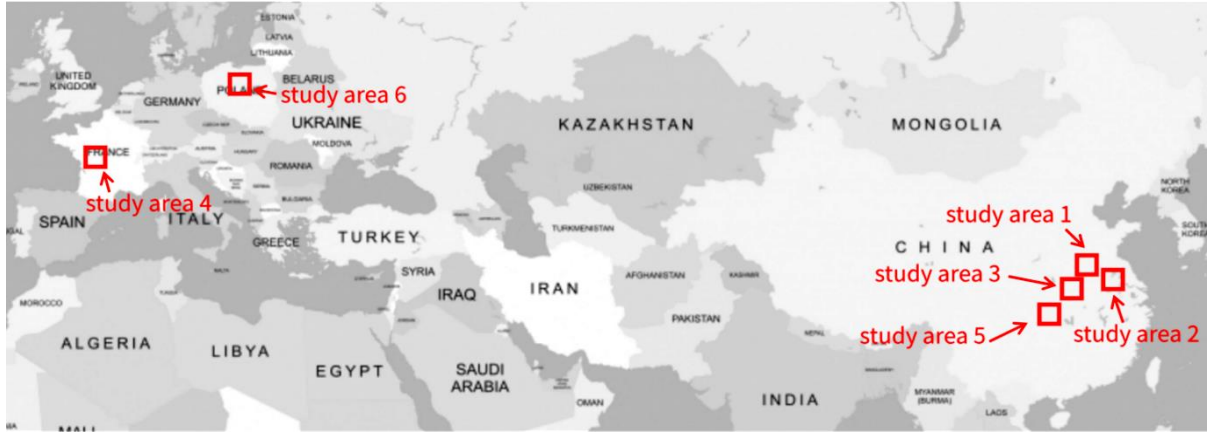

**Figure 4.** Study areas in China and Europe

**Table 2.** Detailed description of experimental data

| Area type | No. | Lat/Lon range | SAR Date Size(pixels) | Optical Date Size(pixels) | Image Characteristic |
|---|---|---|---|---|---|
| Flat | Study 1 | 33.35°N 34.34°N 115.91°E 117.11°E | 2018/10/06 10980×10980 | 2018/10/12 10980×10980 | Images locate in the Midwest of the North China Plain. From the DEM map in Figure 5(a), most of the area is at elevations below 50m except for few highlands in the upper right corner of the area reaching an elevation of about 130m. The maximum elevation difference is less than 120 m, and thereby the area belongs to the flat terrain. |
| Flat | Study 2 | 32.43°N 33.43°N 118.06°E 119.26°E | 2018/04/04 10980×10980 | 2018/03/28 10980×10980 | Images locate in the plain at the junction of Anhui and Jiangsu provinces, China. The terrain fluctuation is small and the elevation is below 160m. From the DEM map in Figure 5(b), although there are some small hills in the southwest corner of the area, more than 75% of the area is flat and the maximum elevation difference is about 150m. Therefore, the area belongs the flat terrain. |
| Hill | Study 3 | 30.60°N 31.62°N 113.09°E 114.26°E | 2018/04/19 10980×10980 | 2018/04/18 10980×10980 | Images locates in the northeast of Hubei province, China. From the DEM map in Figure 5(c), The northwest and northeast of the area are surrounded by many hills. The overall elevation range is between 10m and 250m, and the elevation of two thirds of the area is below 100m. The maximum elevation difference is about 200m, and thereby the area is classified as the hilly terrain. |
| Hill | Study 4 | 46.79°N 47.82°N 0.38°W 1.14°E | 2018/08/12 10980×10980 | 2018/08/02 10980×10980 | Image locates in the Midwest of France, centered at Tours city. From the DEM map in Figure 5(d), the area has many hills and presents an undulating terrain. |

|  |  |  |  |  | The maximum elevation difference reaches 200m, and thereby the area is classified as the hilly terrain. |
|---|---|---|---|---|---|
| Mountain | Study 5 | 28.83°N 29.83°N 110.99°E 112.14°E | 2018/09/27 10980×10980 | 2018/10/05 10980×10980 | Image locates in the western area of Changde City, Hunan Province, China. From the DEM map in Figure 5(e), there is a great contrast in elevation between the east and west of the area. The west is covered by mountains and the elevation is about 700m-800m, while the east is an urban area with elevation ranging from 10m to 150m. The overall terrain is undulating greatly, with the maximum elevation difference of nearly 800 meters. Therefore, the area belongs to the mountain terrain. |
|  | Study 6 | 50.46°N 51.45°N 19.56°E 21.14°E | 2018/02/24 10980×10980 | 2019/02/24 10980×10980 | Image locates in southeastern of Poland, with Kielce city as the center. From the DEM map in Figure 5(f), the mountain range crosses the image area diagonally. Although there are no obvious peaks, but the elevation distribution is very complex, floating around 190-500 meters, with a maximum elevation difference of 300 meters, belonging to the mountain terrain. |

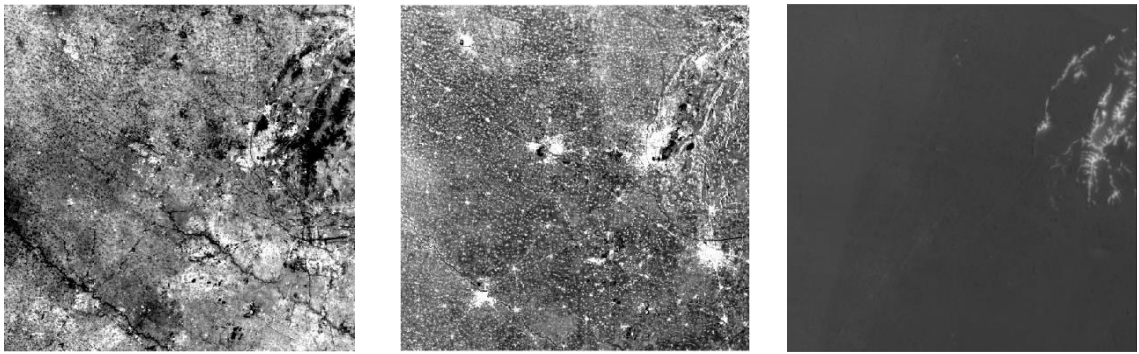

(a)

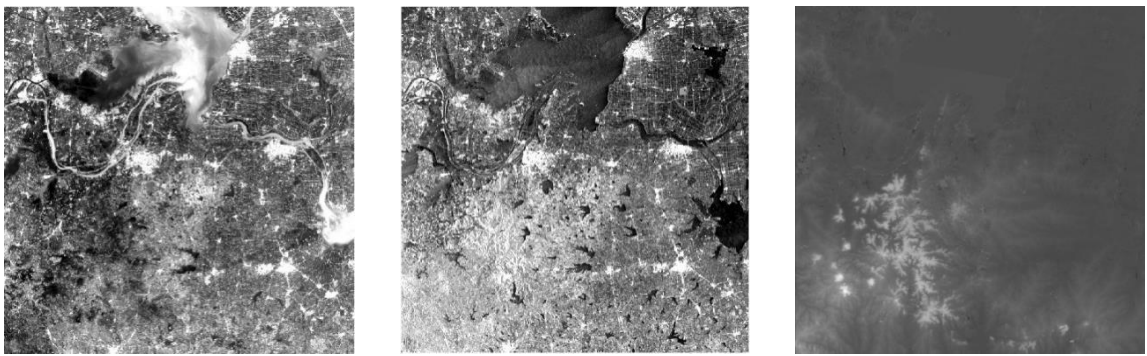

(b)

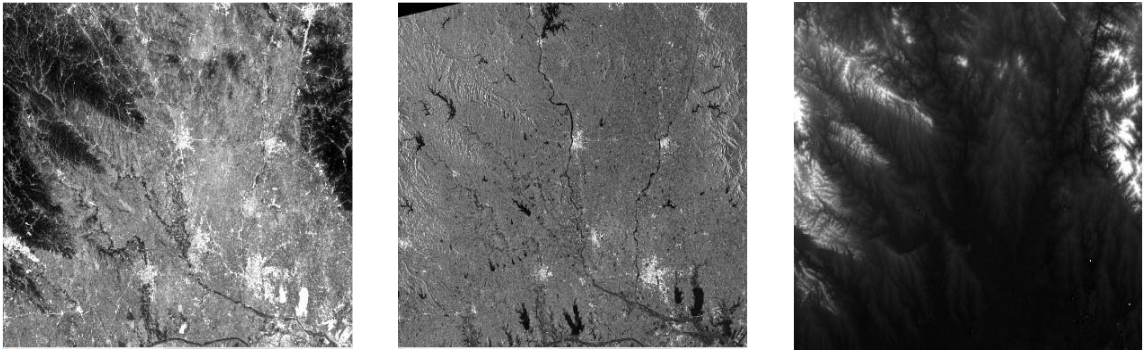

(c)

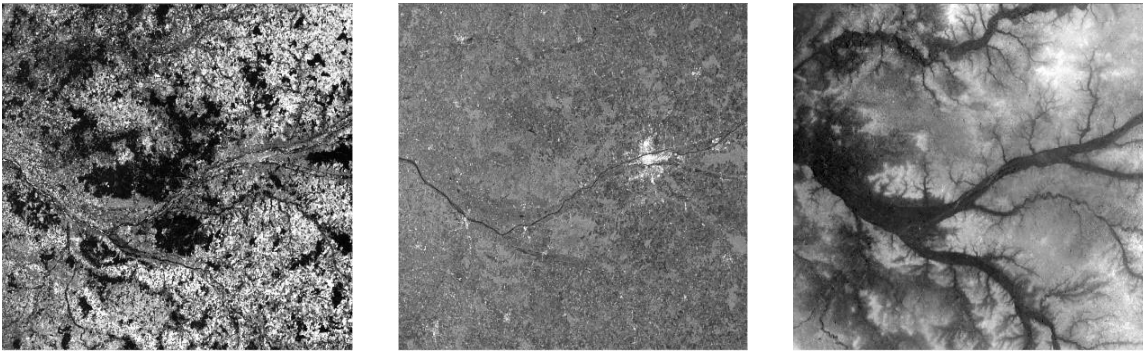

(d)

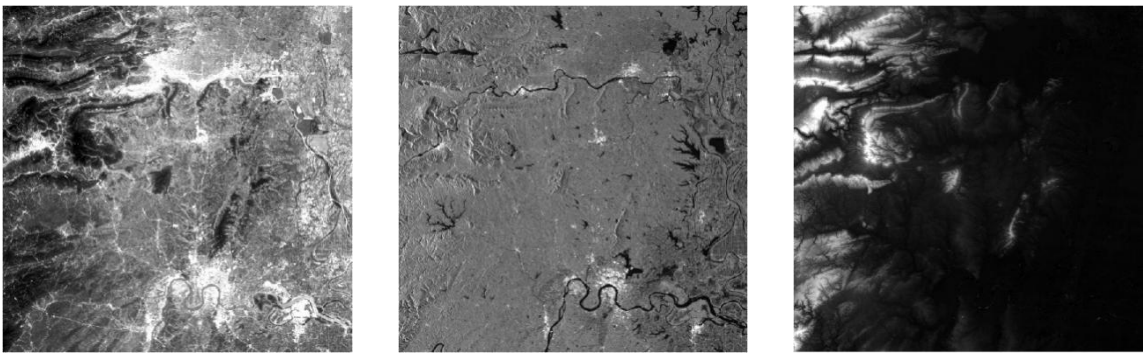

(e)

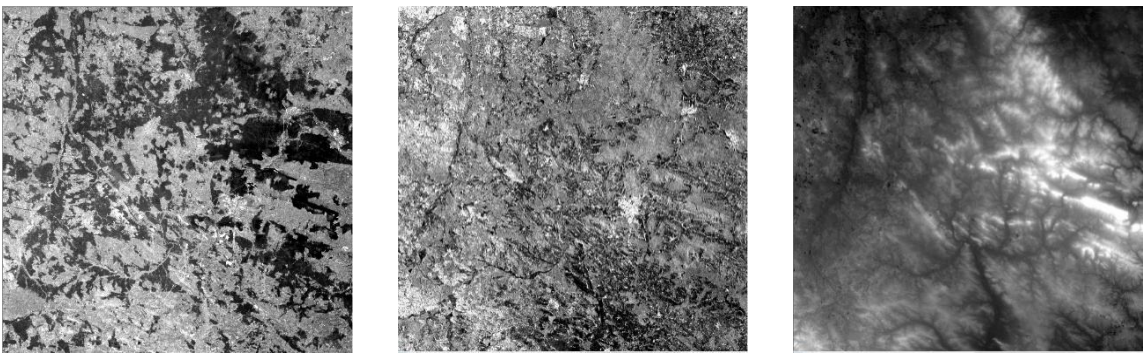

(f)

**Figure 5.** Experimental data including optical images (left), SAR images (middle) and DEMs (right). (**a**) Study area 1. (**b**) Study area 2. (**c**) Study area 3. (**d**) Study area 4. (**e**) Study area 5. (**f**) Study area 6.

*4.2. Image matching*

In the image matching, the optical image is first divided into 20 × 20 girds and one FAST interest point is detected in each grid, reaching a total of 400 interest points. Then their match points in the SAR image is achieved by using CFOG and 3D PC with a template matching manner, where the template window is set to 100 × 100 pixels, and the search window is set to 200 × 200 pixels. Finally, the RANSAC algorithm is employed to remove the outliers to obtain the correspondences. The matching process is performed on a personal computer (PC) with Intel Core i7-10700KF CPU 3.80 GHz. Table 3 shows that the number of the interest points, the correspondences and the run time. It can be found that these image pairs achieve different number of correspondences because of their different image characteristics. Specifically, the images covering flat areas obtain more correspondences than the images covering mountain areas. This may be attributed to that the images of mountain areas have more significant local geometric distortions than these of flat areas. These geometric distortions results in the extracted structural features between the images cannot correspond well, which degrades the matching performance to some degree. In addition, the run time is about 6.6 seconds for all the image pairs, which demonstrates the high computational efficiency of the proposed match method. Overall, our match approach is fast and reliable for the matching of the Sentinel SAR and optical images, and the obtained correspondences for all the tested image pairs are sufficient for their co-registration. To make a fair and reasonable test, it is necessary to use the same number of correspondences to perform the following misregistration measurement and compare the registration accuracy of different transformation models. Accordingly, we select 143 correspondences with least residuals for each image pair in the following tests.

**Table 3.** Matching statistics of all study areas

| No. | Number of extracted interest points | Number of correspondences | Time(s) | Scene type |
|---|---|---|---|---|
| Study 1 | 400 | 318 | 6.59 | Flat area |
| Study 2 | 400 | 259 | 6.52 | Flat area |
| Study 3 | 400 | 304 | 6.57 | Hilly area |
| Study 4 | 400 | 172 | 6.65 | Hilly area |
| Study 5 | 400 | 275 | 6.62 | Mountainous area |
| Study 6 | 400 | 186 | 6.67 | Mountainous area |

*4.3. Misregistration measurement*

According to the obtained 143 correspondences in the previous subsection, we use their mean offsets of match locations to compute the misregistration shifts $\Delta \bar{x}$, $\Delta \bar{y}$ and $\Delta \bar{s}$ (see Equation 15). Figure 6 shows the misregistration shifts of the six pairs of images covering three different terrains. In general, the misregistration shifts presents a large difference for different terrains. The misregistration shifts are kept at about 20-30 pixels in the flat areas and about 20-40 pixels in hilly areas, respectively. While in the mountainous areas with large topographic relief, the misregistration shifts increases sharply to about 50-60 pixels. This is because that Sentinel SAR and optical images have different imaging geometry models, where the SAR sensor acquires data by a side-look imaging way while the optical sensor captures data by a push-broom imaging way. This makes geometric distortions between them more significant with the increase of terrain fluctuation.

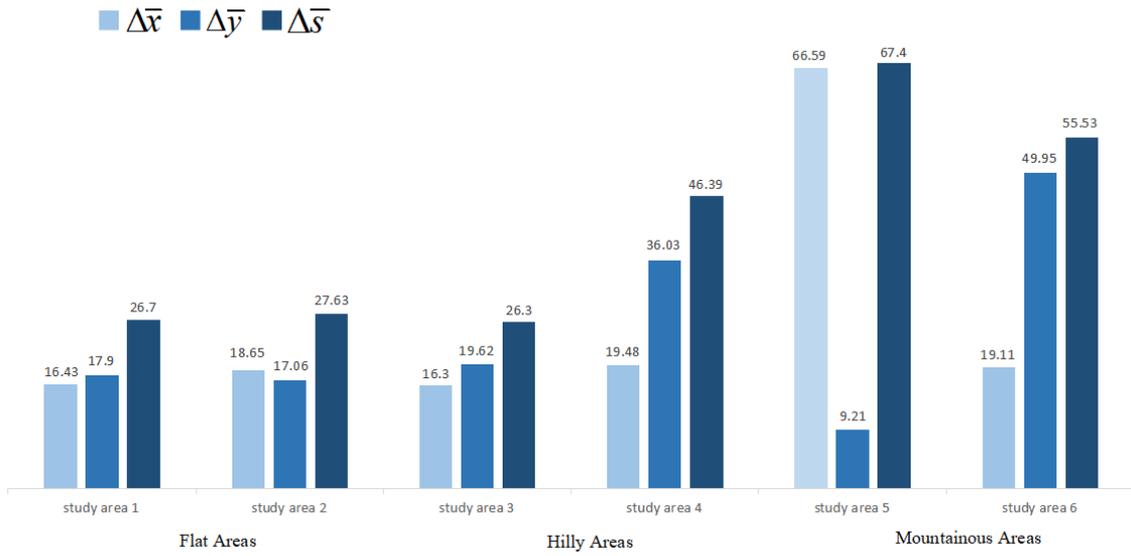

**Figure 6.** Statistics of mean misregistration shifts (units pixels) of correspondences for different terrains

Table 4 gives the statistics of the misregistration shifts for each image pair. We can see that all these images pairs have the large standard deviations (STDs), which means that the misregistration shifts are unevenly distributed across the images, that is, the misregistration shifts are quite different for the correspondences in different regions of each image pair (see Figure 7). The results indicate that these images have significant nonlinear geometric distortions, and different image regions correspond to different misregistration shifts. The main reason for that is the Sentinel-1 SAR images have not undergone the terrain correction.

**Table 4.** Statistics of the misregistration shifts (units pixels) of all the study areas

| Area type | No. | $\Delta \overline{x}$ | $\Delta \overline{y}$ | $\Delta \overline{s}$ | Maximal $\Delta s$ | Minimal $\Delta s$ | STD |
|---|---|---|---|---|---|---|---|
| Flat | Study 1 | 16.43 | 17.90 | 26.70 | 61.06 | 5.39 | 9.84 |
| | Study 2 | 18.65 | 17.06 | 27.63 | 91.24 | 31.62 | 11.79 |
| Hill | Study 3 | 16.30 | 19.62 | 26.30 | 39.30 | 10.70 | 6.05 |
| | Study 4 | 19.48 | 36.03 | 46.39 | 82.15 | 3.61 | 12.06 |
| Mountain | Study 5 | 66.59 | 9.21 | 67.40 | 114.11 | 26.40 | 16.80 |
| | Study 6 | 19.11 | 49.95 | 55.53 | 74.43 | 29.97 | 9.13 |

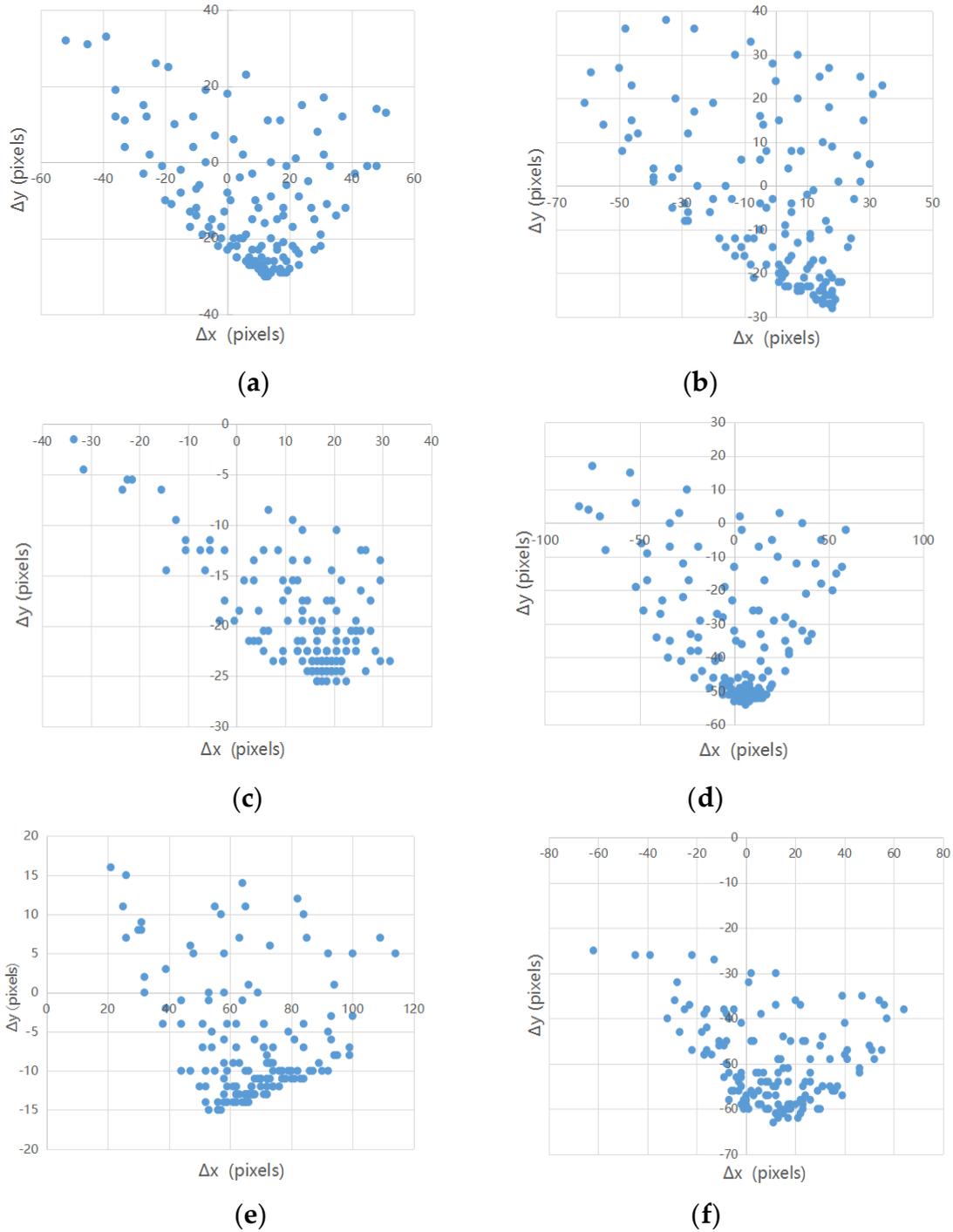

**Figure 7.** Distribution of misregistration shifts (units pixels) $\Delta x$ and $\Delta y$ of different correspondences for all the study areas. (**a**) Study area 1. (**b**) Study area 2. (**c**) Study area 3. (**d**) Study area 4. (**e**) Study area 5. (**f**) Study area 6.

The above results demonstrate that there are evident misregistration shifts between the Sentinel-1 SAR and Sentinel-2 optical images. Accordingly, a further co-registration process for the SAR and optical images is presented in the following.

*4.4. Co-registration accuracy analysis and evaluation*

In this subsection, we will evaluate the registration accuracy of different geometric transformation models for the Sentinel-1 SAR and Sentinel-2 optical images. Firstly, the obtained 143 correspondences (in subsection 4.2) are divided into checkpoints and CPs (see Figure 8), where 48 correspondences are used as checkpoints and the others are used as CPs (their number is variable). Then, we compare the performance of the polynomials (from 1st up to 5st order), the projective transformations (for 10 parameters up to 38 parameters), and the RFMs (from 1st up to 3rd) under different number of CPs (from 25 to 95) and different terrains. Finally, an optimum geometric model is determined for the co-registration of the two types of images. The experimental analysis for different terrains is given in the following.

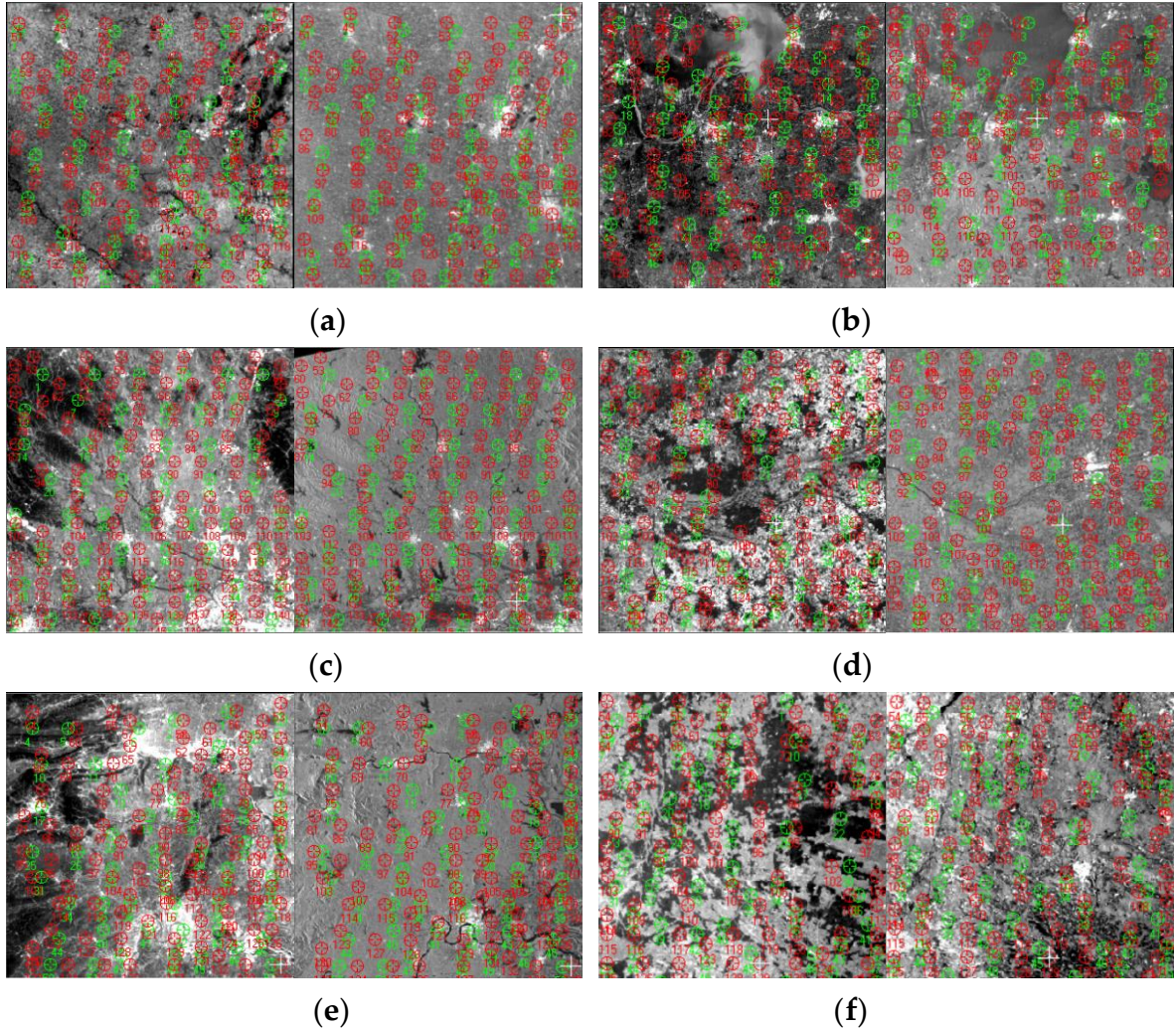

**Figure 8.** CPs (red) and checkpoints (green) for optical (left) and SAR (right) images. (**a**) Study area 1. (**b**) Study area 2. (**c**) Study area 3. (**d**) Study area 4. (**e**) Study area 5. (**f**) Study area 6.

4.4.1. Accuracy analysis of flat areas

Let us first analyze the registration accuracy of the polynomial models. It can be found from Table 5 that the RMSEs of the 1st. Order polynomial are more than 10 pixels for the two flat areas, which are much larger than these of the other polynomials. Accordingly, we mainly compare the registration accuracy of the polynomials (from 2nd. Order to 5th. Order) because their RMSEs are relatively close. Figure 9 shows the checkpoint RMSEs versus the number of CPs for these polynomials. It can be observed that the 2nd. Order and the 3rd.Order polynomials present the stable performance when the number of CPs changes. The 3rd Order polynomial achieves higher registration accuracy than the 2nd. Order polynomial, and its accuracy are kept at about 0.4 pixels for study area 1 and about 0.75 pixels for study area 2, respectively. For the 4th. Order and 5th. Order polynomials, although their accuracy is close to that of the 3rd. Order polynomial under a large number of control points (e.g., more than 80), their performance has a significant fluctuation with the change of the number of CPs. Moreover, they require more CPs for image registration compared with the 3rd. Order polynomial, which will increase computational cost and is not beneficial to practical application. Based on the above analysis, the 3rd. Order polynomial performs better compared with the other polynomials.

Table 5. Checkpoint error statistics (units pixels) of different geometric models in the flat areas, where the number of CPs is 95 and the number of checkpoints is 48

| Geometric model | Study area 1 | | Study area 2 | |
|---|---|---|---|---|
| | Maximum residual | RMSE | Maximum residual | RMSE |
| 1st.Order polynomial | 28.90 | 10.29 | 33.74 | 11.78 |
| 2nd.Order polynomial | 0.78 | 0.40 | 3.42 | 1.03 |
| 3rd.Order polynomial | 0.75 | 0.38 | 3.45 | 0.76 |
| 4th.Order polynomial | 0.92 | 0.44 | 2.98 | 0.74 |
| 5th.Order polynomial | 0.68 | 0.43 | 3.63 | 0.77 |
| 10-parameter projective transformation | 10.19 | 4.38 | 10.11 | 4.53 |
| 22-parameter projective transformation | 1.29 | 0.60 | 3.07 | 1.16 |
| 38-parameter projective transformation | 1.51 | 0.70 | 4.47 | 1.43 |
| 1st.Order RFM (same denominator) | 24.46 | 8.31 | 21.79 | 7.11 |
| 1st.Order RFM (different denominator) | 9.73 | 4.38 | 9.18 | 4.41 |
| 2nd.Order RFM (same denominator) | 7.68 | 1.42 | 11.45 | 3.50 |
| 2nd.Order RFM (different denominator) | 1.94 | 0.68 | 4.67 | 1.46 |
| 3rd.Order RFM (same denominator) | 6.02 | 1.28 | 13.34 | 3.27 |
| 3rd.Order RFM (different denominator) | 1.03 | 0.70 | 3.77 | 1.27 |

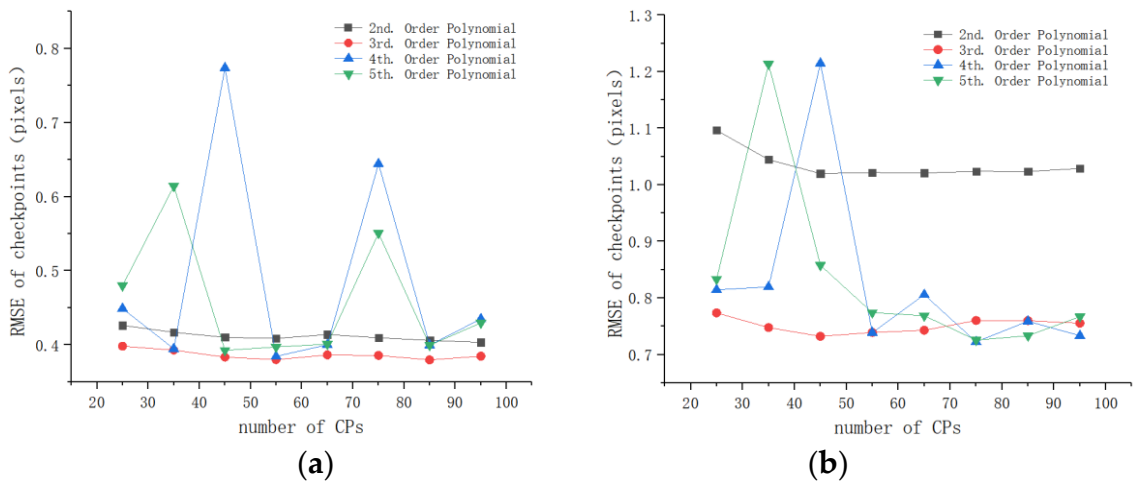

Figure 9. Checkpoint RMSEs of polynomials for the flat areas. (a) Study area 1. (b) Study area 2.

For projective models and RFMs, we can see from Table 5 that the 22-parameter and 38-parameter projective models perform better than the other projective models, and the 2nd.Order and 3rd.Order RFMs (different denominator) achieves higher accuracy than the others. Accordingly, these models are compared with the 3rd. Order polynomial. Figure 10 shows their checkpoint RMSEs versus the number of CPs. It can be clearly observed that the 3rd. Order polynomial achieve the smallest RMSEs under the any amount of CPs among these models. Moreover, the performance of the 3rd. Order polynomial are more stable than that of the other models when the number of CPs changes. These results demonstrate that the 3rd. Order polynomial is the best geometric transformation model for the co-registration of the Sentinel-1 SAR and Sentinel-2 optical images in the flat areas.

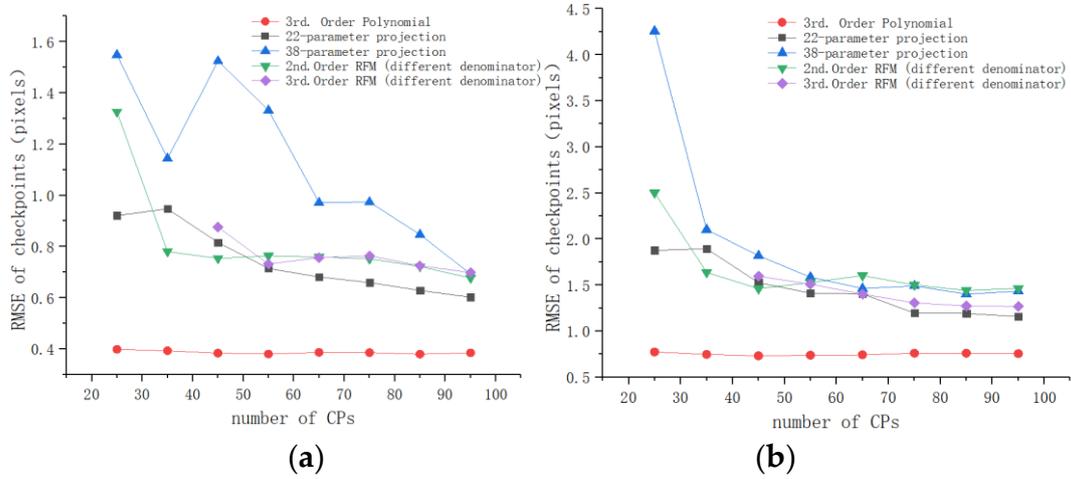

**Figure 10.** Checkpoint RMSEs of the 3rd. Order polynomial and other geometric models for the flat areas. (**a**) Study area 1. (**b**) Study area 2.

4.4.2. Accuracy analysis of hilly areas

The experimental analysis is the similar to that in the flat areas. We also first compare and analyze the registration accuracy of the polynomials. From Table 6, we can see that the 1st Order polynomial performs much worse than the other polynomials. Accordingly, the comparative analysis focuses on the polynomials of 2nd. Order to 5th. Order. Figure 11 shows the checkpoint RMSEs versus the number of CPs for these polynomials. It can be seen that the 2nd. Order polynomial obtains lower accuracy than the other polynomials, while the 3rd. Order, the 4th. Order, and the 5th. Order polynomials achieve the similar accuracy. However, in study area 4, the accuracy of the 4th. Order polynomial has a sharp fluctuation when the number of CPs changes (see Figure 11(**b**)), which shows its instability for image registration. The 3rd. Order and the 5th.Order polynomials present the stable performance and have the same level of accuracy, but the 3rd. Order polynomial requires less CPs and is more computational efficient than the 5th.Order polynomial. Accordingly, the 3rd. Order polynomial is more preferable than the other polynomials.

**Table 6.** Checkpoint error statistics (units pixels) of different geometric models in the hilly area, where the number of CPs is 95 and the number of checkpoints is 48.

| Geometric model | Study Area 3 | | Study Area 4 | |
|---|---|---|---|---|
| | Maximum residual | RMSE | Maximum residual | RMSE |
| 1st.Order polynomial | 29.71 | 12.17 | 32.03 | 12.95 |
| 2nd.Order polynomial | 4.76 | 1.59 | 6.34 | 1.76 |
| 3rd.Order polynomial | 4.07 | 1.35 | 3.27 | 1.46 |
| 4th.Order polynomial | 3.74 | 1.33 | 3.32 | 1.56 |
| 5th.Order polynomial | 3.91 | 1.32 | 3.39 | 1.47 |
| 10-parameter projective transformation | 10.76 | 4.02 | 9.92 | 5.45 |
| 22-parameter projective transformation | 5.79 | 2.11 | 5.28 | 2.47 |
| 38-parameter projective transformation | 4.43 | 1.83 | 6.65 | 2.39 |
| 1st.Order RFM (same denominator) | 24.72 | 7.43 | 25.33 | 8.59 |
| 1st.Order RFM (different denominator) | 8.03 | 3.37 | 11.43 | 5.20 |
| 2nd.Order RFM (same denominator) | 14.24 | 3.82 | 13.12 | 4.86 |
| 2nd.Order RFM (different denominator) | 5.53 | 1.84 | 6.67 | 2.34 |
| 3rd.Order RFM (same denominator) | 8.87 | 2.90 | 8.85 | 4.33 |
| 3rd.Order RFM (different denominator) | 4.80 | 1.64 | 5.46 | 1.86 |

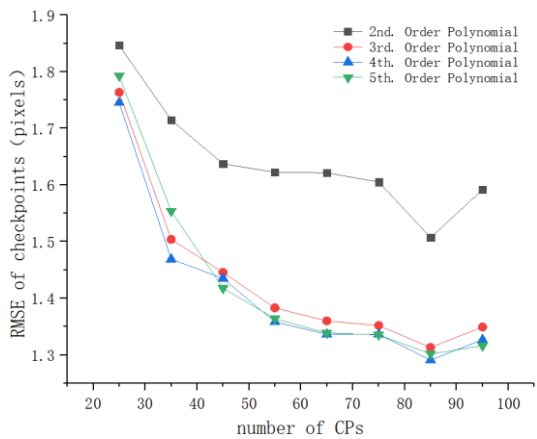
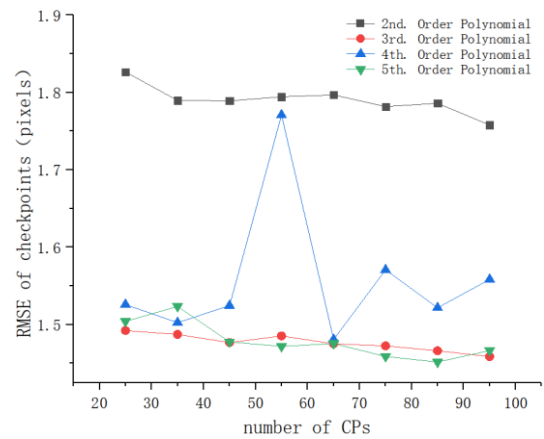

(a)    (b)

**Figure 11.** Checkpoint RMSEs of polynomials for the flat areas. (**a**) Study area 3. (**b**) Study area 4.

For the other geometric models, the ones with higher accuracy are selected to compare with the 3rd. Order polynomial. We can see from Table 6 that these models are the 22-parameter projective transformation, the 38-paremeter projective transformation, the 2nd. Order and the 3rd. Order RFMs (different denominator). Figure 12 shows the checkpoint RMSEs versus the number of CPs for these models. Apparently, the 3rd. Order polynomial achieves a registration accuracy of about 1.5 pixels for both study area 3 and area 4, and outperforms the other models under any mount of CPs. These results confirm that the 3rd Order polynomial is optimal among these models for the co-registration of the Sentinel-1 SAR and Sentinel-2 optical images in the hilly areas.

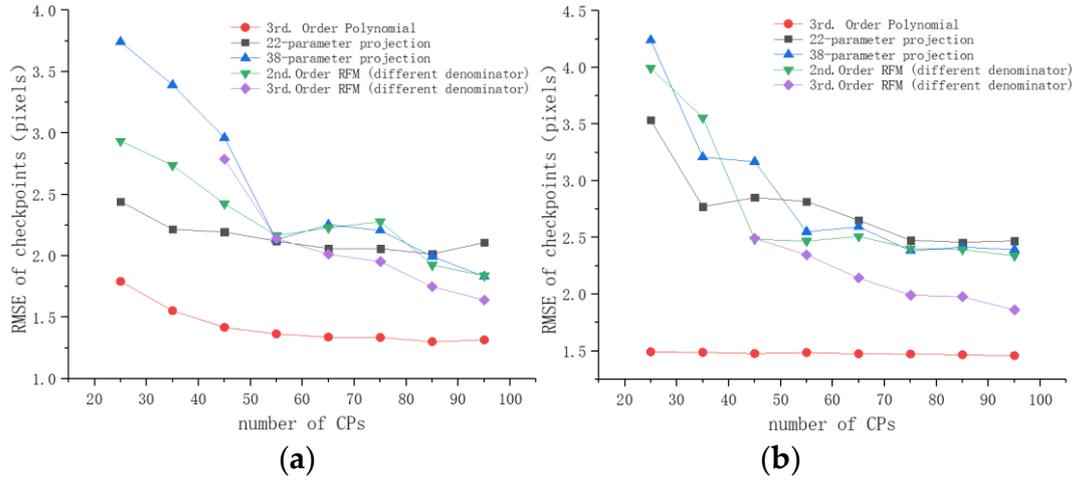

**Figure 12.** Checkpoint RMSEs of the 3rd. Order polynomial and the other geometric models for the hilly areas. (**a**) Study area 3. (**b**) Study area 4.

4.4.3. Accuracy analysis of mountainous areas

Figure 13 shows the checkpoint RMSEs versus the number of CPs for the polynomials. We can see that the 2nd. Order polynomial perform much worse than the other polynomials. The accuracy of the 5th. Order polynomial has a significant fluctuation when the number of control points changes. By comparison, the 3rd. Order and the 4th. Order polynomials have more stable performance and achieves higher accuracy. Compared the 4th. Order polynomial, the performance of the 3rd. Order polynomial is more stable for study areas 6, and it requires less CPs. Accordingly, the 3rd. Order polynomial is more suitable than the other polynomials for the co-registration of these images.

**Table 7.** Checkpoint error statistics (units pixels) of different geometric models in the mountainous areas, where the number of CPs is 95 and the number of checkpoints is 48.

| Geometric model | Study Area 5 | | Study Area 6 | |
|---|---|---|---|---|
| | Maximum residual | RMSE | Maximum residual | RMSE |
| 1st.Order polynomial | 35.55 | 14.41 | 48.58 | 17.94 |
| 2nd.Order polynomial | 5.13 | 2.77 | 13.48 | 3.63 |
| 3rd.Order polynomial | 4.14 | 1.74 | 5.76 | 2.29 |
| 4th.Order polynomial | 4.48 | 1.70 | 5.82 | 2.31 |
| 5th.Order polynomial | 4.33 | 1.65 | 8.32 | 2.83 |
| 10-parameter projective transformation | 9.98 | 3.43 | 17.47 | 8.48 |
| 22-parameter projective transformation | 7.31 | 2.03 | 7.03 | 3.48 |
| 38-parameter projective transformation | 5.11 | 1.80 | 8.53 | 3.37 |
| 1st.Order RFM (same denominator) | 20.98 | 6.57 | 45.31 | 14.95 |
| 1st.Order RFM (different denominator) | 10.26 | 2.95 | 18.17 | 7.59 |
| 2nd.Order RFM (same denominator) | 5.11 | 2.08 | 15.67 | 8.01 |
| 2nd.Order RFM (different denominator) | 4.16 | 1.82 | 7.82 | 3.25 |
| 3rd.Order RFM (same denominator) | 5.94 | 2.14 | 14.30 | 6.68 |
| 3rd.Order RFM (different denominator) | 4.38 | 1.84 | 4.33 | 2.44 |

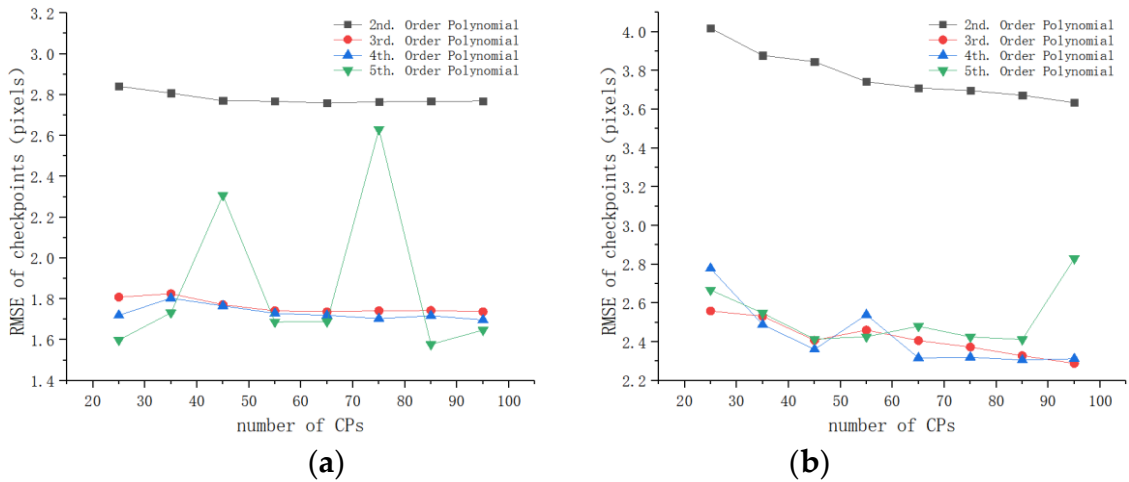

**Figure 13.** Checkpoint RMSEs of the polynomials for the mountainous areas. (**a**) Study area 5. (**b**) Study area 6.

The same as the analysis scheme in the flat areas and hilly areas, we choose other several models with higher accuracy from Table 7, and compare them with the 3rd. Order polynomial. Figure 14 shows the checkpoint RMSEs versus the number of CPs for these models. For study area 5, the accuracy of the 3rd. Order polynomial is about 1.7 pixels, which is much better than that of the other models. In the case of study area 6, the accuracy of the 3rd. RFM model improves with the increase of the number of CPs, and its accuracy is close to that of the 3rd. Order polynomial when the number of CPs reaches 95. However, the 3rd. Order polynomial still yields slightly better accuracy than the 3rd. RFM model. The accuracy is about 2.3 pixels. Furthermore, the 3rd. Order polynomial require less control points compared with the 3rd. RFM model. Therefore, the 3rd. Order polynomial outperforms the other geometric transformation models for the co-registration of the Sentinel-1 SAR and Sentinel-1 optical images in the mountainous areas.

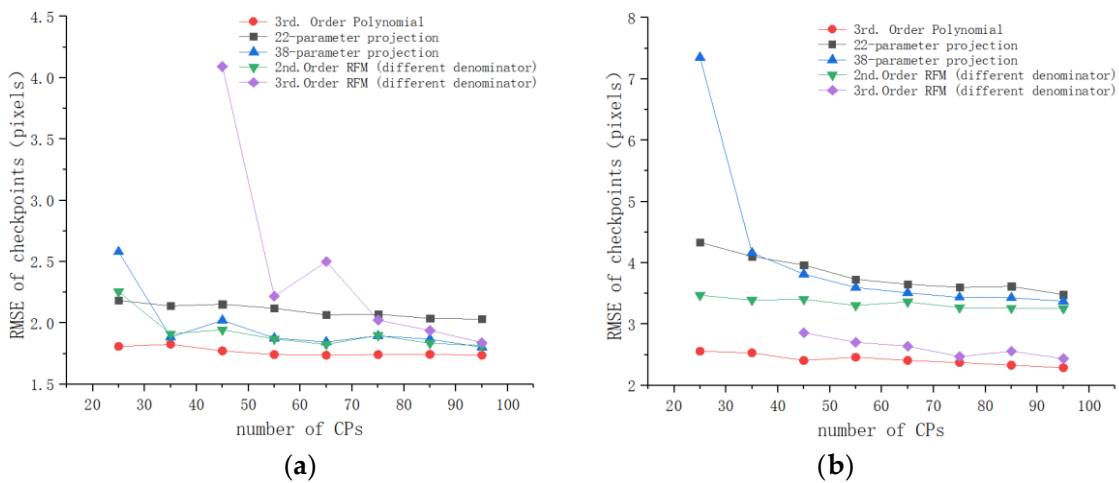

**Figure 14.** Checkpoint RMSEs of the 3rd. Order polynomial and the other geometric models for the Mountainous areas. (**a**) Study area 5. (**b**) Study area 6.

4.4.4. Evaluation of registration results

Based on the above experimental analysis, the 3rd. Order polynomial is the optimal geometric transformation model for the co-registration of the Sentinel-1 SAR and Sentinel-2 optical images, and thereby this model is used to perform the image registration. This process just takes about 3 seconds for one image pair by using our PC. Since the previous image matching process takes about 6.7 seconds, the proposed method costs less than 10 seconds to complete the whole co-registration.

To illustrate the advantage of the proposed method, it is compared with the terrain correction process using the SNAP toolbox. Table 8 gives the registration accuracy and run time of the two methods. We can see that the proposed method outperforms the terrain correction process for most test cases in registration accuracy, especially for the images covering flat and hilly areas. Moreover, the proposed method is almost 100 times faster than the terrain correction process, which is quite beneficial for the response of emergency events such as earthquakes and floods. The registration results is shown in Figure 15. It can be observed that the Sentinel-1 SAR and Sentinel-2 optical images have been aligned well.

**Table 8.** Comparison of registration results of the proposed method and the terrain correction process.

| No. | Proposed method | | Terrain correction | | Scene type |
|---|---|---|---|---|---|
| | RMSE(pixels) | Time(s) | RMSE(pixels) | Time(s) | |
| Study 1 | 0.38 | 9.59 | 0. 92 | 909 | Flat area |
| Study 2 | 0.76 | 9.52 | 1.12 | 892 | Flat area |
| Study 3 | 1.35 | 9.57 | 1.52 | 917 | Hilly area |
| Study 4 | 1.46 | 9.65 | 1.61 | 904 | Hilly area |
| Study 5 | 1.74 | 9.62 | 1.87 | 898 | Mountainous area |
| Study 6 | 2.29 | 9.67 | 2.27 | 919 | Mountainous area |

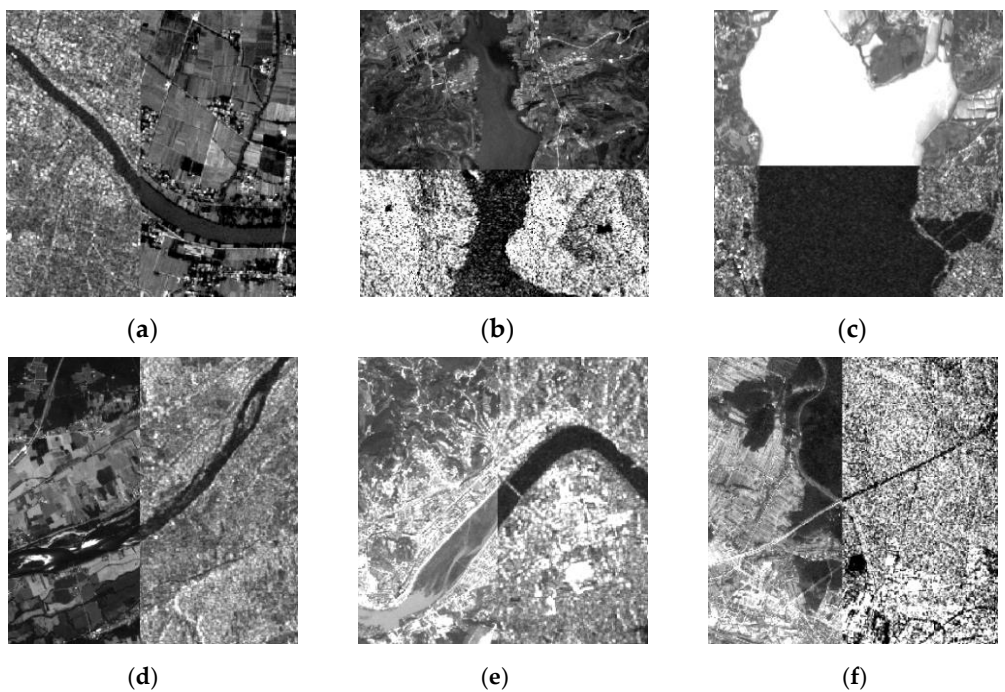

**Figure 15.** Examples of the Sentinel-1 SAR and Sentinel-2 optical image registration. (**a**) Study area 1. (**b**) Study area 2. (**c**) Study area 3. (**d**) Study area 4. (**e**) Study area 5. (**f**) Study area 6.

## 5. Discussions and Conclusions

This paper first proposed a fast and robust block-based match scheme based on structural features and 3D PC for correspondence detection between the Sentinel-1 SAR L1 IW GRD 10m images and the Sentinel-2 optical L1C 10m images. Then the obtained correspondences are used to measure the misregistration shifts of the two types of images and analyse the performance of various geometric transformation models for their co-registration. Finally, an optimal geometric model is determined to improve the registration accuracy.

In the image matching, we apply the various techniques to ensure the matching quality. The block-based scheme aims to make the correspondences evenly distributed over the images, the CFOG structural feature descriptor can handle significant radiometric differences between the images, and 3D PC greatly improves the computational efficiency. The mismatches are removed by using the RANSAC algorithm. Accordingly, the obtained correspondences are reliable for the misregistration measurement and the accuracy evaluation of geometric transformation models.

The misregistration is measured by computing the geometric shifts of correspondences between the two types of images. Six pairs of images covering different terrains are matched and used for the misregistration measurement, as well as for the subsequent co-registration tests. Experimental results show that the misregistration shifts between them are about between 20 and 60 10m pixels, and the shifts vary drastically in different regions of images. Furthermore, the misregistration increases with the increase of topographic relief. Specifically speaking, the misregistration of the flat areas is kept at 20-30 10m pixels, and that of the hilly areas is kept at 20-40 10m pixels, and that of mountainous areas increase to 50-60 10m pixels. Such large of registration shifts result in that the two types of images cannot be effectively integrated for the subsequent remote sensing applications without a precise co-registration. Since only the six pairs of images are used to measure the misregistration shifts, the shifts may be different when using the images located in other areas and acquired at other times. However, this research finds that the misregistration shifts are common between the Sentinel-1 SAR L1 and Sentinel-2 optical L1C images. Moreover, the larger topographic relief is, the larger the misregistration shifts are.

To improve the co-registration accuracy between the Sentinel-1 SAR and Sentinel-2 optical images, we compare the performance of a variety of geometric transformation models including polynomials, projective models, and RFMs. In general, considering some factors such as registration accuracy, required number of CPs, and stability of performance, the 3rd. Order polynomial performs better than the other models, while the projective models don't achieve the satisfactory registration accuracy. For the RFMs, especially the 3rd. Order RFM, although it considers the influence of terrain height and has been widely used for the correction and registration of high-resolution images, its performance is not quite satisfactory for the co-registration of the Sentinel SAR and optical images. The main reason may be that the parameters of the RFMs are calculated by plenty of CPs in the case of high-resolution images [31]. Whereas in our experiments, the number of CPs are limited within 95, which may not be able to provide the exact solution for the RFM parameters. Accordingly, the accuracy of the RFMs could be improved by increasing the number of CPs. In a word, our experiments show that the 3rd. Order polynomial is the optimal geometrical model under a limited number of CPs (e.g., less than 95) for the co-registration of the Sentinel-1 SAR and Sentinel-2 optical images.

Compared with the co-registration process pipeline that the terrain correction is applied on the Sentinel-1 SAR images by using the SNAP toolbox, the proposed method is almost 100 times faster, and obtains higher registration accuracy for most tested image pairs, especially for the images covering flat and hilly areas. Therefore, the proposed method can effectively improve the co-registration for the Sentinel SAR and optical images, and also can meet the requirement of real-time processing. In addition, the registration accuracy of the proposed method varies for different terrains. When using the optimal geometric model (i.e., the 3rd. Order polynomial), The flat areas achieve the highest accuracy that reaches the sub-pixel (probably 0.4 to 0.75 pixels), followed by the hilly areas which achieve an accuracy of about 1.5 pixels, whereas the mountainous areas have the lowest accuracy which is between 1.7 and 2.3 pixels. These results illustrate that the registration accuracy decrease with the increase of topographic relief. This is because large topographic relief increases local geometric distortions between images, and the 3rd. Order polynomial only can approximately fit these distortions. To track this problem, a possible solution is to perform a true ortho-rectification using the rigorous physical models and orbit ephemeris parameters of Sentinel-1 and Sentinel-2 satellites. Meanwhile, a high resolution DEM data also should be used in this process. These work will be carried out in future research. Moreover, we also will test some non-rigid geometric transformation models such as the piecewise-linear and thin-plate-spline functions for the co-registration of the Sentinel-1 SAR and Sentinel-2 optical images.

**Author Contributions:** Yuanxin Ye and Chao Yang developed the method and wrote the manuscript. Bai Zhu, Liang Zhou, Youquan He and Huarong Jia designed and carried out the experiments. All the authors analyzed the results and improved the manuscript.

**Acknowledgments:** This paper is supported by the National Natural Science Foundation of China (No.41971281).

**Conflicts of Interest:** The authors declare no conflict of interest